\documentclass[10pt, aps, prc, onecolumn, nofootinbib, 
preprintnumbers, superscriptaddress, amsmath,floatfix, amssymb, longbibliography]{revtex4-1}

\usepackage{graphicx,amsmath,amssymb,bm}
\usepackage{amsfonts}
\usepackage[utf8]{inputenc}
\usepackage{verbatim}
\usepackage{float}\usepackage[labelfont={small},font=small,subrefformat=parens,caption=false]{subfig}
\usepackage{cancel}
\usepackage{multirow}
\usepackage{array}
\usepackage{xparse}
\usepackage{xspace}

\usepackage{bigstrut}

\usepackage{color}
\usepackage{dsfont}

\newcommand{\pr}{{\rm pr}}
\newcommand{\matA}{{A}}
\newcommand{\be}{\begin{equation}}
\newcommand{\ee}{\end{equation}}

\graphicspath{{Figures/}}

\begin{document}

\title{Does Bayesian Model Averaging improve polynomial extrapolations? Two toy problems as tests}

\author{M.~A.~Connell}
\email{mc959616@ohio.edu} 
\affiliation{Department of Physics and Astronomy, Ohio University, Athens, OH 45701, USA}

\author{I.~Billig}
\affiliation{Department of Physics and Astronomy, Ohio University, Athens, OH 45701, USA}

\author{D.~R.~Phillips}
\email{phillid1@ohio.edu} 
\affiliation{Department of Physics and Astronomy and Institute of Nuclear and Particle Physics, Ohio University, Athens, OH 45701, USA}

\date{\today}
\begin{abstract}

We assess the accuracy of Bayesian polynomial extrapolations from small parameter values, $x$, to large values of $x$. We consider a set of polynomials of fixed order, intended as a proxy for a fixed-order effective field theory (EFT) description of data. We employ Bayesian Model Averaging (BMA) to combine results from different order polynomials (EFT orders). Our study considers two ``toy problems'' where the underlying function used to generate data sets is known. We use Bayesian parameter estimation to extract the polynomial coefficients that describe these data at low $x$. A ``naturalness'' prior is imposed on the coefficients, so that they are $\mathcal{O}(1)$. We Bayesian-Model-Average different polynomial degrees by weighting each according to its Bayesian evidence and compare the predictive performance of this Bayesian Model Average with that of the individual polynomials. The credibility intervals on the BMA forecast have the stated coverage properties more consistently than does the highest evidence polynomial, though BMA does not necessarily outperform every polynomial.
\end{abstract}      
\maketitle

\section{Introduction}

Effective field theory (EFT) organizes the low-energy interactions of particles in a systematic way~\cite{Weinberg:1978kz,Gasser:1983yg,Kaplan:1995uv,Scherer:2012xha}. It does this by separating low-momentum physics from high-momentum physics, following the guiding principle that the high-momentum physics has limited impact at low momentum. The EFT contains an infinite string of interactions induced by the unresolved high-momentum physics. But, for any given reaction, these interactions can be organized as a series in powers of the momentum at which the reaction takes place, $k$, divided by the high-momentum scale, $\Lambda$.
The higher-order terms in the series then have successively smaller impact at low energies. The effects of physics at the scale $\Lambda$ on observables are parameterized via
Low-Energy Constants (or LECs) that are the coefficients occurring at different orders $(k/\Lambda)^n$ of the EFT expansion. If the EFT is to be phenomenologically useful, the values of the LECs have to be determined. This then allows the observable to be evaluated at momenta, $k$, where it has not been measured. The resulting description of the observable is valid for $k$ below $\Lambda$, so $\Lambda$ is the breakdown scale of the EFT. 

Bayesian methods have found extensive use in nuclear-physics EFTs~\cite{Schindler:2009,Furnstahl:2015,Griesshammer:2015ahu,
Wesolowski:2016,Melendez:2017,Wesolowski:2018lzj,Melendez:2019,Drischler:2020hwi,Drischler:2020yad,Premarathna:2019tup,Filin:2019eoe,Filin:2020tcs,PhysRevLett.126.092501,Wesolowski:2021cni}. In a Bayesian EFT framework prior information on LECs is combined with data to yield the posterior probability distribution of EFT coefficients. In the case of small toy  data sets, Bayesian methods yield better estimated LECs than do least-squares fitting techniques; prior knowledge of the LECs' size prevents overfitting of data ~\cite{Schindler:2009,Wesolowski:2016}.
The systematic character of the EFT expansion also allows the
error made when the EFT result is truncated at a finite order to be quantified~\cite{Furnstahl:2015,Melendez:2017}, and the correlations between those errors at different kinematic points to be assessed~\cite{Melendez:2019,Drischler:2020hwi,Drischler:2020yad}.

Once LECs have been extracted from low-momentum data the EFT can be used to extrapolate from that data set to higher-momentum ``target points''. (Extrapolating to zero energy and interpolating data are related problems and we do not study them here.) But, if several different orders in the EFT give plausible fits to the available data, which order should we use as the extrapolant? In particular, as higher values of the reaction momentum are considered, the different EFT orders may predict quite different extrapolants, so this ambiguity as to which is best introduces model uncertainty into the extrapolation. 
As ~\cite{Hoeting:1999} states in a slightly different context: ``ambiguity about model selection should dilute information about effect sizes and predictions.''

This concern about the correct order to use for an EFT extrapolation can be addressed using Bayesian Model Averaging (BMA)~\cite{Hoeting:1999,Was00}. 
To perform BMA we compute the evidence that each model describes the data, and create a ``mixed model'' that is the average of the models' posterior probability density function (pdf), weighted by these evidences. BMA thus provides a way to assess model uncertainty, since different models that provide equally good fits to the data will all contribute to the BMA prediction for an observable. If the models disagree on the observable's predicted value the resulting BMA pdf will spread over more values than will the prediction of any one model. BMA thus yields a prediction that encompasses all the plausible predictions of the models that are mixed. BMA is one example of the more general technique of Bayesian Model Mixing. (See Ref.~\cite{Phillips:2020dmw} for further discussion of the ways in which BMM is more general than BMA.)

BMA has found application in many fields~\cite{Fragoso:2015}. In nuclear physics it has recently been used to analyze nuclear mass models and improve predictions for the location of the drip line~\cite{Kejzlar:2020vla,Neufcourt:2019sle,Neufcourt:2019qvd} as well as to extract transport coefficients from heavy-ion collision data in a way that accounts for uncertainty in the ``particlization'' model~\cite{Everett:2020yty}. Jay and Neil used both mock and actual data to demonstrate that it is a useful way to account for different choices of minimum and maximum time separation when fitting a two-point lattice QCD correlator~\cite{Jay:2020jkz}. But Ref.~\cite{Jay:2020jkz} is the only controlled test of BMA in a nuclear-physics context that we are aware of.

In this paper, we make such a test using two toy problems that mimic efforts to build an extrapolant based on EFT~\cite{Schindler:2009,Wesolowski:2016,Drischler:2020hwi,Drischler:2020yad}. 
Let $f$ represent a generic observable that is computed in an EFT. We take $f$ to have been rescaled so that it is dimensionless. Suppose $f$ depends on the momentum $k$ at which it is measured, and define $x=k/\Lambda$. $f$ is then a function of $x$, with the finite-order EFT result becoming a less accurate approximant to the true result the larger $x$ gets. We denote the LECs by $a_i$ and write
\begin{equation} \label{eq:EFTform}
f_M(x)=a_0+a_1x+a_2x^2+\ldots=\sum_{i=0}^M a_i x^i,
\end{equation}
where we also have indicated the order of the polynomial (EFT) by the subscript $M$. (We do not consider observables with non-analytic pieces here, since the coefficients of non-analytic terms in the EFT expansion are typically determined in other processes or by lower-order LECs~\cite{Weinberg:1978kz,Gasser:1983yg,Scherer:2012xha}. The function $f_M(x)$ could be considered the order-$M$ piece of the EFT prediction after known non-analytic effects have been removed from the data.)

In each toy model we generate ``toy problem data'' from two different underlying functions $g_1$ and $g_2$. The data include some ``experimental" uncertainty. We calculate posteriors for the coefficients for several polynomial fits of different degree $M$ to these data. The resulting fits are then mixed via BMA to generate a mixed model. We test how BMA performs as it extrapolates away from the domain where the pseudodata set lies, by computing the result obtained for the extrapolation at a new ``target" point $x_t$. 
Since we chose the underlying function(s) that generates the data, we can check whether the mixed model is a statistically superior extrapolant than any of the degree-$M$ polynomials. Predictive performance is assessed via a ``Credibility Interval Diagnostic''~\cite{Bastos:2009} (i.e., by calculating ``Empirical Coverage Probabilities'') that quantifies the success of the extrapolation in forecasting the value of the underlying function at the target point. 

In an `$\mathcal{M}$-closed' problem the underlying function is part of the set of models. In the $\mathcal{M}$-closed context and with enough data BMA will eventually converge to weights of 1 for the correct model and 0 for all other models, i.e., the mixing procedure will hone in on the correct model~\cite{Chib:2016}. 
In our case a finite polynomial cannot perfectly replicate either $g$. 
This is because both $g_1$ and $g_2$ are outside the model set of finite polynomials. Both problems are therefore what Ref.~\cite{Hoeting:1999} terms `$\mathcal{M}$-open', since the set of models being mixed does not incorporate the true model.  
In the $\mathcal{M}$-open context there is no guarantee that BMA will give a useful result~\cite{BernardoSmith1994}. 

In Sec.~\ref{sec:twotoys} we define the two underlying functions we use. They have very different properties and, although both represent situations that are formally $\mathcal{M}$-open, the functions' overall behavior influences the degree to which BMA succeeds. Function $g_1$ diverges at $x=1$, and all its Taylor series coefficients are positive. Both our polynomial and BMA extrapolations of data generated by this function fail, providing an explicit demonstration that---in spite of its consideration of model uncertainty---BMA is not a panacea for extrapolation of data. Function $g_2$ converges to 0 as $x$ increases, and the Taylor series coefficients are alternately positive and negative. This is a much easier extrapolation problem. We use $g_2$ to study the nuances of BMA, measuring BMA predictive performance in the case when several non-mixed models are already well-suited for extrapolating.  Sec.~\ref{sec:twotoys} also explains how we generate pseudodata sets from $g_1$ and $g_2$.

In Section~\ref{sec:form}, we discuss the formal aspects of our Bayesian framework. We specify the LEC prior as a Gaussian of width $\sigma_a$ and also specify the prior for $\sigma_a$. We then explain how the posterior for the LECs is computed and how it can be used
to generate an observable posterior $\pr(f|D,M,\sigma_a)$. The resulting $f_M(x_t)$ is then the degree-$M$ polynomial forecast for the underlying function at a target point $x_t$ not in the initial data set, given a particular value of $\sigma_a$. We deal with uncertainty in $\sigma_a$ by marginalizing over it~\cite{Wesolowski:2016} and so we explain how to obtain the $\sigma_a$ posterior that is input to that marginalization.
Then, we address the evidence, or the $M$ posterior, and introduce the formulae for BMA that yield a mixed  posterior $\pr(f(x_t)|D)$ at a target point $x_t$.

Section~\ref{sec:result} contains our results. We begin by demonstrating LEC extraction in our analysis for the calculation of $f_M$, before giving examples of the $\sigma_a$ and $M$ posteriors obtained from pseudodata.
We perform BMA and demonstrate how mixed models behave when extrapolating from the different underlying functions and at different distances for the target point $x_t$. 

In Sec.~\ref{sec:cid} we define the Credibility Interval Diagnostic (CID), which we use to check if the performance of an extrapolation agrees with the credibility intervals obtained from its posterior pdf. We present CID plots to compare the predictive performance of non-mixed models to our mixed model. We dissect how the properties of the mixed models are defined by the non-mixed models that compose it, and we quantify the differences between our underlying functions with these CID plots.

In Section~\ref{sec:end}, we review the differences we see between mixed and non-mixed models. We also establish the patterns for successes and failures of BMA, based on which underlying function, $g$, was used, and how extrapolation distance affects the mixed models' ability to forecast. We propose a set of circumstances in which BMA can be effective.

\section{The two toy-model functions and pseudo-data generation therefrom}

\label{sec:twotoys}

We use two different functions to generate sets of  pseudodata. Different ``EFT orders" (polynomials of degree $M$) are then used to approximate these underlying functions.
Previous Bayesian EFT research~\cite{Schindler:2009,Wesolowski:2016} employed these underlying functions, which were chosen for their analytic structure, and consequent properties of their Taylor expansions. Crucially, both functions have Taylor series coefficients of $\mathcal{O}(1)$, a property we expect in the LECs of a well-behaved EFT expansion.

\subsection{The functions $g_1$ and $g_2$} 

We first use~\cite{Schindler:2009}
\begin{equation} \label{eq:g1}
g_1(x) = \left(\frac{1}{2}+\tan\left(\frac{\pi}{2}x\right)\right)^2 = 0.25 + 1.57 x + 2.47 x^2 + 1.29 x^3 + \ldots \, .
\end{equation}
This function has a pole at $x=1$, so as we approach $x=1$ the underlying function diverges and outpaces any finite polynomial extrapolation. Relatedly, each coefficient is positive in the Taylor expansion of this function, so finite-order polynomial approximants, with LECs estimated at small positive $x$, tend to underestimate the value of the function as $x \rightarrow 1$.

The second function~\cite{Wesolowski:2016} is more docile. It is 
\begin{equation} \label{eq:g2}
g_2(x) = \left(\frac{1.3}{1.3+x}\right)^2 = 1 - 1.54 x + 1.78 x^2 - 1.87 x^3 + 1.75 x^4 + \ldots \, .
\end{equation}
The pole at $x=-1.3$ prevents data at target points $x_t > 0$ from flying away from polynomial extrapolations, as happens with $g_1(x)$. For $g_2(x)$, the coefficients of its Taylor expansion alternate, so a finite polynomial expansion may underestimate or overestimate $g_2(x)$.
Polynomial modeling and model mixing for $g_2(x)$ turn out to be successful more frequently than they are with $g_1(x)$, as $g_1(x)$'s pole impedes accurate finite polynomial extrapolation to target points near $x=1$. Thus $g_2$ helps us determine how BMA improves simple extrapolations, while $g_1$ acts as a stress case where we can check whether BMA provides any benefit.

\subsection{Pseudodata details}

Our underlying functions are used to generate sets of 10 data points $\{(x_i,d_i)\}$, uniformly spaced in the interval $x \in [0,1/\pi]$. The data point $d_i$ is randomly moved away from the underlying function by an error that is distributed as a Gaussian with mean zero and standard deviation $\sigma_i=0.05*g(x_i)$. This is then represented by the error bars on each data point. 

Various pseudodata sets resulting from this procedure are used to extract unique LECs: each data set produces slightly different central values and uncertainties, although once $M$ is large enough the LEC determinations from different data sets are consistent~\cite{Wesolowski:2016}. 

When we test our models' extrapolation to a target point, we generate a ``target data point'' or ``validation data'' $d_t$ at $x_t$ from the underlying distribution $g(x_t)$ via the same method as our pseudodata, i.e., by applying a 5\% normally distributed error to $g$ when sampling this point. We return to this in Subsection ~\ref{ssec:ciddefine}.

We refer to each set of 10 data points as a different ``pseudodata set''. To reproduce the results found in Refs.~\cite{Schindler:2009} and ~\cite{Wesolowski:2016}, we use the same data set employed for LEC extraction in this research (and provided in those works). But for all other results, we are using unique sets of pseudodata. In this way we confirm that our results are generally applicable for these underlying functions and are not dependent on the specifics of any particular pseudodata set.

\section{Bayesian Formalism} 
\label{sec:form}

\subsection{The LEC prior} 

\label{ssec:LECpriors}

In our Bayesian approach the coefficients within each polynomial are given a prior distribution, with this prior chosen in accordance with the physical principles encoded in an EFT. The LECs in an EFT expansion account for the effect of short-distance physics on the observable. LECs that are too large or small can occur if that short-distance physics is fine tuned and exerts an unusual degree of influence on the observable. They can also arise in observables where particular suppressions lead to the vanishing of lower-order effects, thereby rendering higher-order physics more important than would otherwise be the case. But typically, if the breakdown scale has been correctly identified, and the kinematic quantity ($k$ in this case) expressed in units of the breakdown scale, the coefficients of $x \equiv k/\Lambda$ in the EFT expansion should be of order 1, a property called  ``naturalness''. The breakdown in $x$ of the Taylor-series expansion of the two underlying functions is at $|x|=1$ ($g_1(x)$) and $|x|=1.3$ ($g_2(x)$), respectively. And the resulting Taylor-series coefficients on the right-hand side of Eqs.~(\ref{eq:g1}) and (\ref{eq:g2}) are indeed $\mathcal{O}(1)$. We now build this connection between the radius of convergence of the EFT and the size of the LECs into our data analysis, by imposing a prior on the LECs that represents our knowledge that they are $\mathcal{O}(1)$.

We take the prior for the coefficients of our model to be a multidimensional normal distribution of mean $0$ and variance $\sigma_a^2$. Explicitly
\begin{equation} \label{eq:LECprior}
\pr(a_0,a_1\ldots|M,\sigma_a) = \left(\frac{1}{\sigma_a\sqrt{2\pi}}\right)^{M+1}\exp\left(-\frac{a_0^2+a_1^2+\ldots}{2\sigma_a^2}\right).
\end{equation}
$\sigma_a$ is then a hyperparameter that describes how wide we think the distribution can be while still encoding naturalness. 
Smaller values of $\sigma_a$ indicate preference for smaller LECs, and would greatly limit our coefficients. In contrast, as $\sigma_a$ approaches $\infty$, the prior becomes infinitely wide and flat, and the Bayesian parameter estimation process becomes a traditional $\chi^2$-minimization problem (least-squares regression) in this limit. 
The inclusion of a prior with $\sigma_a \approx 2$--$5$ prevents overfitting to limited data that can lead to unnatural LECs~\cite{Schindler:2009,Wesolowski:2016}. But both too small and too large values of $\sigma_a$ are unhelpful, and there is no obvious single value we should choose. Ultimately we address this uncertainty by specifying a prior distribution on $\sigma_a$ values and marginalizing over it.

\subsection{The $\sigma_a$ prior}

\label{ssec:sigmaaprior}

Ref.~\cite{Schindler:2009} took $\sigma_a$ as a constant hyperparameter, and then checked robustness of results against that choice. This makes analysis easier, and since the data will primarily define the LECs, this convention has merit. This would be equivalent to using a $\delta$-function prior on $\sigma_a$. In this section we discuss a less informative $\sigma_a$ prior, which nevertheless is straightforward to compute with. Ultimately the posterior probability of $\sigma_a$, given its prior and our pseudodata, will quantitatively specify what ``naturalness'' means for the LECs in our problem(s). 

Before beginning we note  that our priors on $M$ and $\sigma_a$ are uncorrelated. When specifying all our priors, we will alternate between writing $\pr(M,\sigma_a)$ and $\pr(M)\pr(\sigma_a)$; these are identical.

The maximum entropy prior for positive scale parameters is the Jeffreys prior:
\begin{equation} \label{eq:jeff}
\pr(\sigma_a) \propto \frac{1}{\sigma_a}.
\end{equation}
This prior significantly favors small values of $\sigma_a$, diverging if we choose to extend our analysis to $\sigma_a=0$. In consequence it weights values of $\sigma_a$ near 0 more strongly than is implied by standard definitions of naturalness in an EFT. 
Because of this we choose a prior that is less biased towards small $\sigma_a$ values. The prior is conjugate to the LECs' Gaussian prior $\pr(\vec{a}|M,\sigma_a)$, which simplifies the analysis. 
It is~\cite{Melendez:2019}:
\begin{equation} \label{eq:invchi}
\pr(\sigma_a|\nu_0,\tau_0) = 2 \frac{(\nu_0 \tau_0^2 /2)^{\nu_0/2}}{\Gamma(\nu_0/2)\sigma_a^{1+\nu_0}} \exp\left(-\frac{\nu_0 \tau_0^2}{2\sigma_a^2}\right).
\end{equation}
The hyperparameters, $\nu_0$ and $\tau_0$ determine the peak of the prior and how strongly it extends towards large $\sigma_a^2$.
For $\nu_0=0$ Eq.~(\ref{eq:invchi})  reduces to Eq.~(\ref{eq:jeff}), the Jeffreys prior. 
We employ the Jeffreys prior as well as priors of the form (\ref{eq:invchi}) with $\nu_0=\tau_0= 1, 1.5, 2$. 
Figure 12 in Ref.~\cite{Maris:2020} demonstrates this distribution for these values of $\nu_0$ and $\tau_0$.
Here we present our probabilities as $\pr(\sigma_a)$ rather than the inverse-$\chi^2$ distribution of $\pr(\sigma_a^2)$. The relationship between the pdfs is: 
\begin{equation}
    \pr(\sigma_a)\propto\sigma_a\pr(\sigma_a^2).
\end{equation}
This only changes the presentation of the probability, rather than the actual analysis from Ref.~\cite{Melendez:2019}.

We examined the influence this prior choice has on the extrapolated value and uncertainty at the target point $x_t$. The prior had no impact on the overall success or failure of extrapolation, as assessed through the CID plots we introduce in Sec.~\ref{sec:cid}. Larger values of $\nu_0$ and $\tau_0$ reduce the posterior probability for very low values of $\sigma_a$, and so widen the error bar on the extrapolated value slightly. We choose a representative prior of $\nu_0=\tau_0=1.5$ for $\sigma_a$ for all of our analyses and plots, unless otherwise specified.

\subsection{Fixed-$M$ posteriors for the LECs $\vec{a}$ and the observable $f(x)$}
\label{ssec:LECfpost}
We use Bayes' Theorem to find the $M+1$ dimensional LEC posterior $\pr(\vec{a}|D ,M,\sigma_a)$, accounting for both the prior and pseudodata $D$: 
\begin{equation} \label{eq:LECpost}
\pr(\vec{a}|D,M,\sigma_a)=\frac{\pr(D|\vec{a},M,\sigma_a)\pr(\vec{a}|M,\sigma_a)}{\pr(D|M,\sigma_a)}.
\end{equation}
Here $\pr(D|\vec{a},M,\sigma_a)$ is the likelihood and $\pr(\vec{a}|M,\sigma_a)$ is the prior (\ref{eq:LECprior}), determined before data fitting. The denominator $\pr(D|M,\sigma_a)$ is related to the evidence, and is a constant in terms of $\vec{a}$. Therefore in parameter estimation it acts as a normalization constant.

We take $\pr(D|\vec{a},M,\sigma_a)$ to be the standard (uncorrelated) likelihood $\propto \exp(-\chi^2/2)$, and therefore have:
\begin{equation} \label{eq:LECpost2}
\pr(\vec{a}|D,M,\sigma_a) \propto \exp\left(-\frac{\chi^2}{2}\right)\prod_{i=0}^{M} \exp\left(-\frac{a_i^2}{2\sigma_a^2}\right).
\end{equation}
The standard $\chi^2$ is thus ``augmented'' by the prior term, and so we define
\begin{equation} \label{Chiaug}
\chi_{aug}^2=\sum_{j=1}^N \frac{(d_j - f_M(x_j))^2}{\sigma_j^2} +\sum_{i=0}^{M} \frac{a_i^2}{\sigma_a^2},
\end{equation}
where $N$ is the number of pseudodata points, and have
\begin{equation} \label{eq:LECpost3}
\pr(\vec{a}|D,M,\sigma_a) \propto \exp\left(-\frac{\chi_{aug}^2}{2}\right). \end{equation}

This $M+1$-dimensional Gaussian posterior for $\vec{a}$ has its maximum value at the point
\begin{equation}
    \vec{a}_{0}=\matA_{aug}^{-1}\vec{b}.
\end{equation}
Here we've defined $\matA$ as the ``design matrix'' and $\vec{b}$ as a vector of length $M+1$:
\begin{eqnarray}
A_{i,j}&=&\sum^N_{k=1} \frac{x_k^{i+j}}{\sigma_k^2} ; \;\;\; i,j=0,\ldots,M,\\
b_{i}&=&\sum^N_{k=1} \frac{d_k  x_k^{i}}{\sigma_k^2} ; \;\;\; i=0,\ldots,M,
\end{eqnarray}
and the prior ``augments" $\begin{matrix}A\end{matrix}$ to:
\begin{equation}
\matA_{aug}=\matA+\sigma_a^{-2}\mathbb{I}.
\end{equation}
When the $\chi^2_{aug}$ of the polynomial model is evaluated with the parameter values $\vec{a}_0$ we get the minimum augmented $\chi^2$, $\chi^2_{aug,min}$. 

Refs.~\cite{Schindler:2009,Wesolowski:2016,Melendez:2019}  focused on the LEC posterior, as extracting these coefficients was the goal of these studies. Our goal here is to analyze the extrapolation of data, i.e., evaluate the observable $f_M(x)$, as well as to assess the suitability of BMA for the $f$-posterior evaluation. 

Since the LEC posterior for a single model is an $M+1$-Gaussian distribution, and this is a linear model, the $f_{M,\sigma_a}$ posterior is also Gaussian, and can be defined by its mean and variance~\cite{KlcoThesis:2015}:
\begin{equation}
    \pr(f(x)|D,M,\sigma_a) = \frac{1}{\sqrt{2\pi}\sigma_{f_{M,\sigma_a}}} \exp\left(-\frac{\left(f(x)-\overline{f}_{M,\sigma_a}(x)\right)^2}{2\sigma^2_{f_{M,\sigma_a}}}\right).
    \label{eq:posteriorpredictive}
\end{equation}
Here the mean $\overline{f}_{M,\sigma_a}(x)$ and variance $\sigma_{f_{M,\sigma_a}}^2$ are given by
\begin{equation} \label{eq:fmean}
\overline{f}_{M,\sigma_a}(x)=\sum_{i=0}^{M} \overline{a}_{i} x^i,
\end{equation}
\begin{equation} \label{eq:fvar}
\sigma_{f_{M,\sigma_a}}^2=\sum_{i,j=0}^{M} \sigma_{i,j}^{2} x^{i+j},
\end{equation}
where $\sigma^2_{i,j}$ is the $(i,j)$th element of the covariance matrix between $a_i$ and $a_j$, $\matA^{-1}_{aug}$.

These allow us to calculate the pdf of each polynomial (specific $M$) model at the target point. No model will perfectly predict the data at every target point, so overall we are interested in how well the predictions' error bars describe the underlying distribution at the target.

\subsection{Marginalizing over $\sigma_a$}

So far, our pdfs for the LECs and $f_{M,\sigma_a}$ require a specified value of $\sigma_a$. But we can marginalize over $\sigma_a$ at any place in our analysis. For example, we could marginalize over it in the 
LEC posterior $\pr(\vec{a}|D,M,\sigma_a)$ to form $\pr(\vec{a}|D,M)$ and extract LECs that do not rely on a particular choice of $\sigma_a$. This is useful if LEC extraction is the ultimate goal---see Ref.~\cite{Wesolowski:2016}.

Here our focus is the observable $f$, so we perform marginalization on the $f$ posterior, changing $\pr(f(x)|D,M,\sigma_a)$ into $\pr(f(x)|D,M)$ to obtain a prediction for $f_M(x)$ that does not depend on $\sigma_a$:
\begin{eqnarray}
    \pr(f|D,M)&=&\int \pr(f|D,M,\sigma_a) \pr(\sigma_a|D,M) d\sigma_a \nonumber \\
    &=&\int  \frac{1}{\sqrt{2\pi}\sigma_{f_{M,\sigma_a}}} \exp\left(-\frac{\left(f-\overline{f}_{M,\sigma_a}\right)^2}{2\sigma^2_{f_{M,\sigma_a}}}\right) \pr(\sigma_a|D,M) d\sigma_a.
    \label{eq:margsigmaa}
\end{eqnarray}
This posterior does not have a Gaussian distribution, unless we were to take $\pr(\sigma_a|D,M)$ to be a $\delta$ function. We will address the complications of a non-Gaussian $f_M(x)$ and $f(x)$ in Subsection~\ref{ssec:BMApost}.

We must calculate the posterior distribution of $\sigma_a$, 
which we approach in our usual Bayesian fashion:
\begin{equation}
\pr(\sigma_a|D,M)=\frac{\pr(D|M,\sigma_a)\pr(M)\pr(\sigma_a)}{\pr(D,M)}\propto \pr(D|M,\sigma_a)\pr(\sigma_a),
\end{equation}
where the last step follows because we take a uniform prior for $\pr(M)$ (see Subsection~\ref{ssec:evidence}), so that factor can be incorporated into our normalization constant.

We have defined $\pr(\sigma_a)$, and must give a formula for $\pr(D|M,\sigma_a)$. This can be obtained from the LEC posterior we already have. $\pr(D|M,\sigma_a)$ is the marginal likelihood for a given model $M$ and a particular value of $\sigma_a$. It takes no account of the specific value of $\vec{a}$ we use, and so we marginalize over the LECs:
\begin{equation} \label{eq:intpost}
\pr(D|M,\sigma_a)=\int d\vec{a} \, \pr(D|\vec{a},M,\sigma_a)\pr(\vec{a}|M,\sigma_a).
\end{equation}

Following the logic that led to Eq.~(\ref{eq:LECpost3}) we find that the integrand is proportional to $\exp(-\chi^2_{aug}(\vec{a})/2)$, and it is through the augmented $\chi^2$ that the dependence on $\vec{a}$ enters. But 
 $\chi_{aug}^2$ is quadratic in $\vec{a}$~\cite{Wesolowski:2016}, so we can use the Laplace approximation to evaluate this integral exactly. Following the same steps as Ref.~\cite{Sivia:1996}, but for our Gaussian prior, we then obtain
\begin{equation} \label{eq:finalDev1}
\pr(D|M,\sigma_a) \propto \frac{1}{(\sigma_a)^{M+1}}\sqrt{\frac{1}{{\det(\matA_{aug})}}} \exp\left(-\frac{\chi_{aug,min}^2}{2}\right),
\end{equation}
where the pre-factor comes from the normalization of the LEC prior, $\pr(\vec{a}|M,\sigma_a)$, and must be retained since it is $\sigma_a$ and $M$ dependent. 
Multiplying by the prior (\ref{eq:invchi}) on $\sigma_a$ then yields the posterior of $\sigma_a$ as
\begin{equation} \label{eq:postsig}
\pr(\sigma_a|D,M) \propto  \frac{1}{(\sigma_a)^{M+1}}\sqrt{\frac{1}{{\det(\matA_{aug})}}} \exp\left(-\frac{\chi_{aug,min}^2}{2}\right)\left(\frac{1}{\sigma_a^{1+\nu_0}}\right)\exp\left(\frac{-\nu_0 \tau_0^2}{2\sigma_a^2}\right).
\end{equation}
With $\pr(\sigma_a|D,M)$ in hand, we perform the integration in Eq.~(\ref{eq:margsigmaa}) numerically. 

We can also use the conjugate prior to obtain a semi-analytic form for $\pr(\sigma_a|D,M)$. 
We begin by reducing $\chi^2_{aug,min}$ to a quadratic form (for details, see Appendix~\ref{appendix1}):
\begin{equation} \label{eq:diagchi}
\chi^2_{aug,min} = \vec{\alpha}_{0}^T\frac{1}{\matA^{-1}+\sigma^2_a \mathbb{I}}\vec{\alpha_0} + \chi^2_{min} = \sum_{i=0}^M \frac{(\sum_{j=0}^M O_{i,j} {\alpha_0}_{j})^2}{\Delta_i^{-1}+\sigma^2_a} + \chi^2_{min},
\end{equation}
where $\vec{\alpha_0}$ is the vector of parameters that minimizes the (unaugmented) $\chi^2$, and the design matrix has been diagonalized by $\matA=O^{T}\Delta O$, with $\Delta_i$ the $i$th eigenvalue of $\matA$.

By dividing the $\chi^2_{aug,min}$ as in  Eq. ~(\ref{eq:diagchi}), noting that $\chi^2_{min}$ does not depend on $\sigma_a$, and using the prior on $\sigma_a$ defined by the inverse-$\chi^2$ [Eq.~(\ref{eq:invchi})], we can write the $\sigma_a$ posterior as
\begin{equation} \label{eq:postsig2}
\pr(\sigma_a|D,M) \propto  \frac{1}{(\sigma_a)^{M+2+\nu_0}}\sqrt{\frac{1}{{\det(\matA_{aug})}}} \exp\left(\frac{-\nu_0\tau_0^2}{2\sigma_a^2}\right)\prod_{i=0}^M \exp\left(-\frac{1}{2}\frac{\sum_{j=0}^M (O_{i,j} {\alpha_0}_j)^2}{\Delta_i^{-1}+\sigma^2_a}\right).
\end{equation}

Note that this distribution has the $1/\sigma_a$ factors as well as $\exp(-1/\sigma_a^2)$ behavior we expect in an inverse-$\chi^2$ distribution. We compare this form with  posteriors obtained numerically directly from Eq.~(\ref{eq:postsig}) in Sec.~\ref{subsec:sigmaaresults} below.

\subsection{The evidence}
\label{ssec:evidence}

One obvious question in an EFT calculation is which EFT order (i.e., value of $M$) best fits the data and adheres to the LEC prior. In a Bayesian framework the quantitative answer to this question is  the relative probability for each model given the data, regardless of which coefficients $a_i$ have been chosen within the model: $\pr(Model|Data)$. This ``model evidence'' was computed for the polynomial models considered here in Ref.~\cite{Wesolowski:2016}.  Evidence is only defined relative to the set of models calculated: a meaningful number for each $\pr(M|D)$ is only obtained after all models within the set have been fit to the data, and the overall set normalized, cf. Eq.~(\ref{eq:evidencenorm}). 
It is sometimes assumed that the non-mixed model with the highest evidence is the best model to use for extrapolation, an assumption which we critique in Section~\ref{sec:cid}. 

Through Bayes' theorem we can calculate the model evidence for our case as in ~\cite{Hoeting:1999}:
\begin{equation} \label{eq:Msigpost}
\pr(M|D)=\int d\sigma_a \frac{\pr(D|M,\sigma_a)\pr(M,\sigma_a)}{\pr(D)}
\propto \int d\sigma_a \pr(D|M,\sigma_a)\pr(M)\pr(\sigma_a).
\end{equation}
Here, since the denominator is independent of $M$ and $\sigma_a$, we can ignore it and normalize the results for different $M$  after exhausting our set of models.

Now we must define the prior on $M$. Since we have no prior knowledge about the appropriate polynomial degree, we define a uniform prior for the $M_{max}+1$ different polynomial degrees considered, with the largest model containing LECs $a_0,a_1\ldots a_{M_{max}}$
\begin{equation} \label{eq:Mprior}
\pr(M)=\begin{cases} 
      \frac{1}{M_{max}+1} & 0 \leq M \leq M_{max}\\
      0 & {\rm Otherwise}.
   \end{cases}
\end{equation}
This prior indicates that we know nothing about which degree is best, prior to any data fitting. It thus fulfills the principle of maximum entropy~\cite{Sivia:1996}.

Substituting into Eq.~(\ref{eq:Msigpost}) from Eq.~(\ref{eq:finalDev1}) and employing the inverse-$\chi^2$ prior for $\sigma_a$ reduces the evidence to a one-dimensional integral that we can compute numerically---just as we did for the $\sigma_a$ integral needed to obtain the posterior pdf for $f$, Eq.~(\ref{eq:margsigmaa}):
\begin{equation} \label{eq:Msigpost2}
\pr(M|D)\propto \int d\sigma_a \frac{1}{(\sigma_a)^{\nu_0+M+2}}\sqrt{\frac{1}{{\det(\matA_{aug})}}} \exp\left(-\frac{\chi_{aug,min}^2}{2}\right) \exp\left(-\frac{\nu_0 \tau_0^2}{2 \sigma_a^2}\right).
\end{equation}
The normalization of this model evidence is then obtained using
\begin{equation}
    \sum_{M=0}^{M_{max}} \pr(M|D)=1,
    \label{eq:evidencenorm}
\end{equation}
i.e., by demanding that the set of polynomials considered is exhaustive and mutually exclusive.

\subsection{Model-averaged prediction}
\label{ssec:BMApost}

With posteriors for individual models and calculations of each model's evidence in hand it is now straightforward to form a Bayesian Model Averaged prediction for the function $f$ at a target point $x_t$.

Instead of choosing a single model, BMA addresses model uncertainty in this posterior by mixing the results of multiple models. An averaged or mixed model is a weighted sum of the different models, weighted by the probability that each model is correct. When the models are labeled by a discrete parameter---as is the case here---this is equivalent to marginalizing over that parameter.
Our final mixed-model pdf is defined by the kinematic point $x_t$ where it is evaluated, and so can be written $\pr(f(x_t)|D)$. The posterior is then a function of the target point $x_t$, which may be distant from our pseudodata set. 
\begin{equation} \label{eq:BMAform}
\pr(f(x_t)|D)=\sum_M \pr(f(x_t),M|D) 
=\sum_M \pr(f(x_t)|D,M)\pr(M|D).
\end{equation}

\begin{figure}[h]
\centering
  \subfloat[]
   { \includegraphics[width=0.5\linewidth]{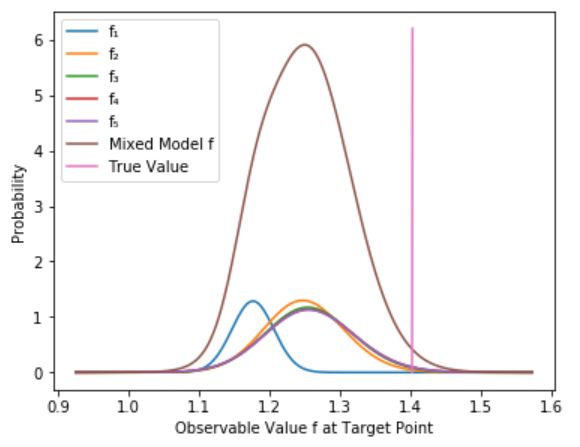}\label{fig:g1BMA}}
\subfloat[]
    {\includegraphics[width=0.5\linewidth]{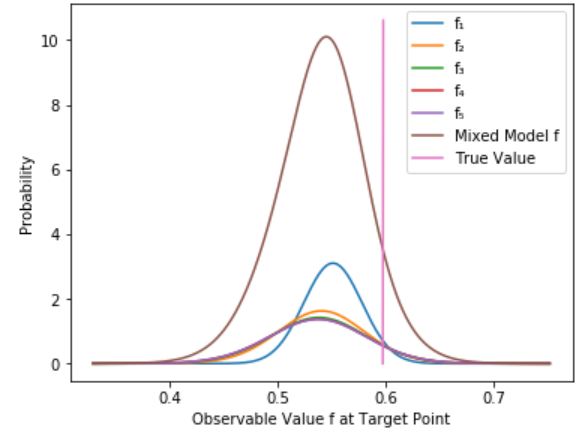}  \label{fig:g2BMA}}
\caption{The probability of particular observable values $f$ at the target point $x_t=1.2/\pi$ in the case of underlying functions $g_1$ (panel (a)) and $g_2$ (panel (b)). Results are shown for extrapolations using $M=1$ (blue), $M=2$ (orange), $M=3$ (green), $M=4$ (red), and $M=5$ (purple). $M=6$ is indistinguishable from $M=5$. 
 The non-mixed $f_M$ distributions are weighted by their evidence. The BMA pdf is shown as the brown line. The true values of $g_1(x_t$ and $g_2(x_t)$ are indicated by the pink line. }
\end{figure}

The sum in Eq.~(\ref{eq:BMAform}) can be visualized by showing each non-mixed model with a height modified by its evidence, see Figs.~\ref{fig:g1BMA} and \ref{fig:g2BMA} for two examples involving extrapolation from a toy data set to $x_t=1.2/\pi$. In each case the final BMA posterior pdf incorporates features of the pdf of the high-evidence models $M=1$ (blue) and $M=2$ (orange). In Fig.~\ref{fig:g2BMA} the $M=1$ result has high evidence but is over-confident. Mixing in the higher-order polynomials produces a better calibrated result. In the case of $g_1$ (Fig.~\ref{fig:g1BMA}) several models have roughly equal evidence. BMA broadens the pdf, and accounts for model uncertainty within this set of models. It yields a finite---albeit still rather small---pdf at the true value. But, because of the behavior of $g_1$, all finite-order polynomials underpredict the true value; no amount of mixing within this set can fix that. We shall return to this point below.

The BMA mean can be found by integrating over $f$
\begin{equation} \label{eq:BMAmean}
\overline{f}=\int f \, \pr(f|D) \, df
=\sum_M \overline{f}_{M} \pr(M|D)
\end{equation}
and the variance is
\begin{equation} \label{eq:BMAvar}
Var(f|D)=\sum_M \pr(M|D) (Var(f|D,M)+\overline{f}_{M}^2)  -\overline{f}^2.
\end{equation}

But a (weighted) mixture of Gaussian pdfs is not itself Gaussian. The $\sigma_a$ marginalization and the averaging over $M$ means the final pdf we obtain is not  Gaussian.
The mean and variance of the BMA distribution thus do not define the distribution as they do for a Gaussian. 
Even though the mean still usually describes where the peak(s) of the BMA distribution is and the variance describes the spread away from the mean, 
the distribution is sometimes multimodal, cf. Fig.~\ref{fig:g1ex2}. Crucially,  it generically has wider tails than a single Gaussian distribution, since BMA combines the tails of multiple models with different means. The translation from variance to credibility intervals is therefore different than it is for a Gaussian. 

\subsection{Model-averaged LECs}

We could extract LECs $a_i$ through BMA as well.
\begin{equation} \label{eq:BMALECmean}
\overline{a}_{i}=\int a_i\pr(a_i|D)da_i
=\sum_M \int \pr(M,\sigma_a|D)\overline{a}_{i,M,\sigma_a} d\sigma_a,
\end{equation}
where we define $\overline{a}_{i,M,\sigma_a}=0$ if $i > M$.
The LEC variance (remembering that the mixed LEC posterior is not Gaussian) is
\begin{equation} \label{eq:BMALECvar}
Var(a_i|D)=\sum_M \int \pr(M,\sigma_a|D) (\sigma_{a_{i,M,\sigma_a}}^2+\overline{a}_{i,M,\sigma_a}^2) d\sigma_a -\overline{a}_i^2.
\end{equation}
Substitution of 
the BMA posterior of $\vec{a}$ into the polynomial returns the same observable pdf $\pr(f(x_t)|D)$ as is found by performing BMA on $f$ itself.

\section{Results}
\label{sec:result}

\subsection{LEC and $f_{M,\sigma_a}$ Posteriors} 

We first present results for individual models, i.e., at fixed values of $M$. At each value of $M$ we extract the posterior for the LECs $a_0, a_1, \ldots, a_M$. We tabulate these results up to $M=4$ in Table ~\ref{tab:g1andg2}. These results were obtained from the
data sets described in Sec.~\ref{sec:twotoys}. The Gaussian naturalness prior on the LECs of Sec.~\ref{ssec:LECpriors} was employed with a $\delta$-function prior on $\sigma_a$ when extracting these LECs. 
This is the same data-generation procedure and prior choice as was used in \textcite{Schindler:2009} and \textcite{Wesolowski:2016}. We reproduce the LEC results in those works when we employ the same values of $\sigma_a$. In Table~\ref{tab:g1andg2} we display two sample sets of results for $\sigma_a=1$ (upper set of results) and $\sigma_a=5$ (lower set of results).
\begin{table}[h]
\begin{center}
 \begin{tabular}{|c | c |c |c |c |c|} 
 \hline
 \multicolumn{6}{|c|}{LEC means and 1$\sigma$ errors for $g_1$, $\sigma_a=1$} \\
 \hline
 $M$ & $a_0$ & $a_1$ & $a_2$ & $a_3$ & $a_4$\\ [0.5ex] 
\hline
 $0$ & $0.456 \pm 0.008$  & $ $  & $ $  & $ $  & $ $\\ 
 \hline
$1$ & $0.190 \pm 0.014$  & $2.513 \pm 0.104$  & $ $  & $ $  & $ $ \\ 
 \hline
$2$ & $0.223 \pm 0.017$  & $1.807 \pm 0.244$  & $2.450 \pm 0.765$  & $ $ & $ $ \\ 
 \hline
$3$ & $0.225 \pm 0.017$  & $1.786 \pm 0.245$  & $2.282 \pm 0.785$  & $0.900 \pm 0.958$  & $ $ \\ 
 \hline
$4$ & $0.226 \pm 0.017$  & $1.785 \pm 0.245$  & $2.265 \pm 0.788$  & $0.891 \pm 0.958$  & $0.284 \pm 0.994$\\ 
 \hline
 \end{tabular}
 \vskip 1 cm
 \begin{tabular}{|c | c |c |c |c |c|} 
 \hline
  \multicolumn{6}{|c|}{LEC means and 1$\sigma$ errors for $g_2$, $\sigma_a=5$} \\
 \hline
 $M$ & $a_0$ & $a_1$ & $a_2$ & $a_3$ & $a_4$\\ [0.5ex] 
 \hline
 $0$ & $0.769 \pm 0.012$  & $ $  & $ $  & $ $  & $ $\\ 
 \hline
$1$ & $0.961 \pm 0.030$  & $0.984 \pm 0.140$  & $ $  & $ $  & $ $ \\ 
 \hline
$2$ & $1.060 \pm 0.052$  & $-2.389 \pm 0.615$  & $3.834 \pm 1.634$  & $ $ & $ $ \\ 
 \hline
$3$ & $1.057 \pm 0.052$  & $-2.289 \pm 0.670$  & $3.042 \pm 2.654$  & $1.651 \pm 4.363$  & $ $ \\ 
 \hline
$4$ & $1.057 \pm 0.053$  & $-2.270 \pm 0.684$  & $2.949 \pm 2.739$  & $1.579 \pm 4.394$  & $0.672 \pm 4.873$\\
 \hline
 \end{tabular}
 \end{center}
 \caption{Tables of LECs for models with $M=0$--$M=4$. The posterior for the LECs is Gaussian, so for each one we list
 $\overline{a}_{i}\pm \sigma_i$. The prior on $\sigma_a$ here is a $\delta$ function, which we only use in these tables.  The upper table uses $g_1$ as the underlying function and we take $\sigma_a=1$, while the lower one uses $g_2$ and $\sigma_a=5$. Note that the mean and variance for each LEC converge with $M$, but that the mean for higher-order LECs converges to 0 while their $\sigma_i$ converges to $\sigma_a$.} 
  \label{tab:g1andg2}
 \end{table}

After finding $\pr(\vec{a}|D,M,\sigma_a)$, we can easily compute $\pr(f(x)|D,M,\sigma_a)$ by Eq.~(\ref{eq:posteriorpredictive}). Equations (\ref{eq:posteriorpredictive})--(\ref{eq:fvar}) tell us that the observable pdf is entirely determined by the LEC posteriors, so the posterior for $f(x_t)$ at fixed $M$ and $\sigma_a$ follows straightforwardly from that of the LECs.

The LECs associated with higher-order polynomial terms are less well-constrained by the data, and they will, as ~\cite{Wesolowski:2016} calls it, `return the prior':
the means approach $0$ and the $\pm1\sigma$ values converge to our prior choice for $\sigma_a$.
This trend of LECs for large $M$ indicates that the higher LECs contribute little to the mean $\overline{f}_{M,\sigma_a}$ and much to the error ${\sigma_f}_{M,\sigma_a}$, see Eqs.~(\ref{eq:fmean}) and (\ref{eq:fvar}).
As $M$ becomes large, adding another LEC $a_{M+1}$ has less effect on the observable mean, but the uncertainty 
associated with this poorly determined parameter causes the observable's variance to increase.
Since the error bars on each LEC become $\pm\sigma_a$ for large $M$, at those values of $M$ the width of $\pr(\vec{a}|D,M,\sigma_a)$ becomes highly $\sigma_a$ dependent. This greatly impacts 
the extrapolation behavior of these models.

The uncertainty as regards the variance of the prediction for $f$ at $x_t$ has two sources: its dependence on $\sigma_a$ and its dependence on $M$. To account for these sources of uncertainty in the prediction we marginalize over $\sigma_a$, and model average over $M$. This manages both our hyperparameter and model selection uncertainties to create a prediction for $f(x_t)$ that does not assume a particular value of either parameter.

\subsection{$\sigma_a$ Posterior}

\label{subsec:sigmaaresults}

We will use Eq.~(\ref{eq:postsig2}) to marginalize over the hyperparameter $\sigma_a$ for each model (value of $M$) This requires $\pr(\sigma_a|D,M)$ as the weighting of the posteriors dependent on $\sigma_a$ inside the integral. 

We approximate the integral in Eq.~(\ref{eq:postsig2})  using the trapezoidal rule. We tested different discretizations of the integral and found that 13 points were sufficient to obtain  approximately $1\%$ accuracy for both $\overline{f}_M$ and $Var(f_M)$.

\begin{figure} [h]
\centering
  \subfloat[]
   { \includegraphics[width=0.5\linewidth]{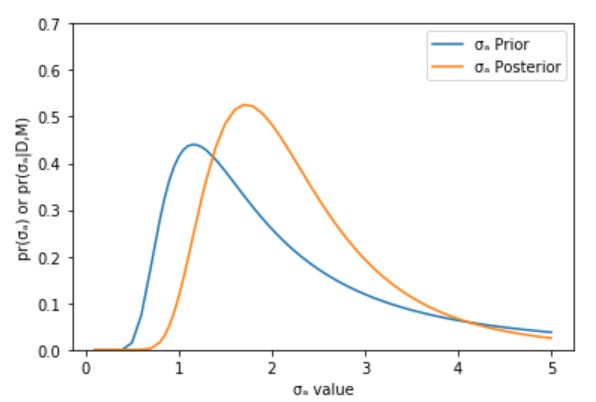}\label{fig:g1sig}}
\subfloat[]
    {\includegraphics[width=0.5\linewidth]{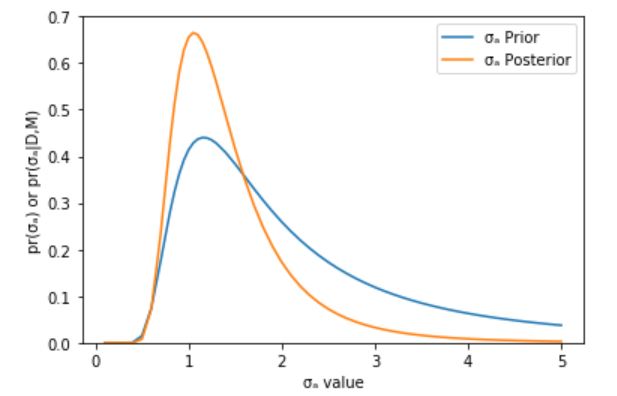}
  \label{fig:g2sig}}
\caption{Normalized prior and posterior pdf of $\sigma_a$ for (a) $g_1$ analysis and (b) $g_2$ analysis. In both cases the polynomial degree is $M=6$. The prior on $\sigma_a^2$ is the same inverse-$\chi^2$ ($\nu_0=\tau_0=1.5$) in both cases.} 
\label{fig:test00}
\end{figure}

We show the $\sigma_a$ posterior in comparison to our $\nu_0=\tau_0=1.5$ inverse-$\chi^2$ prior for the parameter estimation in the $g_1$ (Fig.~\ref{fig:g1sig}) and $g_2$ (Fig.~\ref{fig:g2sig}) cases. In both panels $M=6$. The two panels show that the change in the $\sigma_a$-posterior relative to the prior is different for these two underlying functions. The posterior from $g_1$ has the same width as the prior distribution but its peak is shifted to higher values of $\sigma_a$, while the posterior for $g_2$ is much thinner, with a peak near $\sigma_a=1$. These shifts derive from the properties of the $g$ functions. As we will discuss in more detail below, extrapolation of data is much more difficult for $g_1$  than for $g_2$.
Larger values of $\sigma_a$ restrict our LECs less, allowing bigger values of $a_i$, and the divergence of $g_1$ encourages larger values of these LECs, resulting in a wider $\sigma_a$ posterior. In contrast, $g_2$ can be modeled through smaller LECs, and thus has a thinner peak at small $\sigma_a$.

We already gave the analytic form for the $\sigma_a$ posterior in Eq.~(\ref{eq:postsig2}). Here we reproduce this equation, but now with the individual contributions identified as {\bf [1]}, {\bf [2]}, {\bf [3]}, and {\bf [4]}, so that we can discuss the impact of each piece on the final shape of the posterior.

\begin{equation} \label{eq:sigpostfar}
\pr(\sigma_a|M,D) \propto  \Bigg[\frac{1}{(\sigma_a)^{M+2+\nu_0}}\Bigg]_{\textbf{[1]}}\Bigg[\sqrt{\frac{1}{{\det(\matA_{aug})}}}\Bigg]_{\textbf{[2]}} \Bigg[\exp\left(-\frac{\nu_0\tau_0^2}{2\sigma_a^2}\right)\Bigg]_{\textbf{[3]}} \Bigg[\exp\left\{-\frac{1}{2}\left(\sum_{i=0}^M \frac{\sum_{j=0}^{M} O_{i,j} \overline{a}_{j}}{\Delta_i^{-1}+\sigma^2_a}\right)\right\}\Bigg]_{\textbf{[4]}}
\end{equation}

This formula explains the peak seen in Figs.~\ref{fig:g1sig} and \ref{fig:g2sig}, as well as the rise and fall on either side of it.
For small values of $\sigma_a$, the result is dominated by the term  {\bf [4]} which evaluates $\exp{(-\chi^2_{aug,min}/2)}$. As $\sigma_a \rightarrow 0$ the factor {\bf [2]} approaches 1, and, at least until $\sigma_a^2$ becomes as small as the smallest eigenvalue of the covariance matrix, we have $\exp(-1/\sigma_a^2)$ behavior of the $\sigma_a$ posterior.
As $\sigma_a$ increases after the peak, the term  {\bf [1]} from the $\sigma_a$ prior decreases as $1/\sigma_a^{M + 2 + \nu_0}$, while the determinant factor  {\bf [2]} increases slowly and pieces  {\bf [3]}  and {\bf [4]} converge to a constant. This leads to a power-law falloff for values of $\sigma_a$ above the peak.

If we take the limit where uncertainty on the data becomes very small, the $\Delta_i^{-1}$ term approaches $0$, and the exponential term becomes $\exp(-c/\sigma_a^2)$, causing our posterior to approach an inverse-$\chi^2$ distribution. The posterior thus has a similar shape to the inverse-$\chi^2$ prior but it is not an inverse-$\chi^2$ distribution because the likelihood enters the final result through the presence of the eigenvalues of the covariance matrix $\Delta_i$.

\subsection{Evidence, or the $M$ Posterior}

\begin{figure} [h]
\centering
\begin{subfloat}[]
    {\includegraphics[width=.5\linewidth]{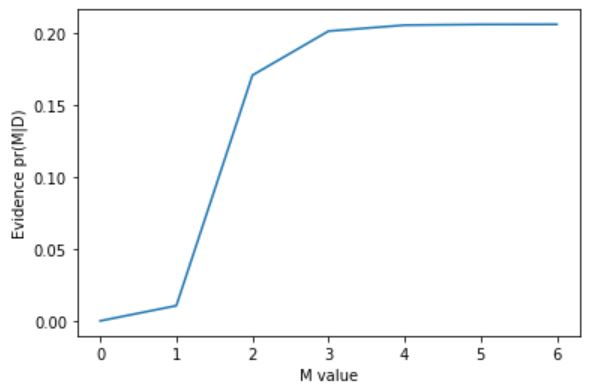}
  \label{fig:g1Mev}}
\end{subfloat}%
\begin{subfloat}[]
  {\includegraphics[width=.5\linewidth]{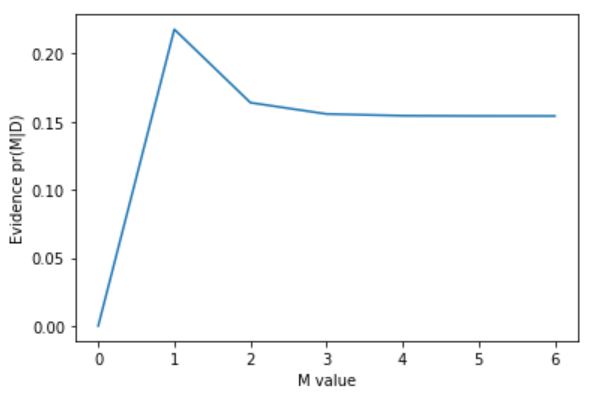}
  \label{fig:g2Mev}}
\end{subfloat}
\caption{The posterior probability of $M$,  the evidence, for $g_1$ (left panel) and $g_2$ (right panel). The prior on $M$ is flat, so all preference for $M$ comes from the data. The $g_1$ case shows no evidence peak in this range.
For $g_2$, a peak occurs at $M=1$, before the curve falls and flattens out afterwards.} 
\label{fig:test0}
\end{figure}

Results for the evidence for different values of $M$ are shown in Figs.~\ref{fig:g1Mev} (pseudodata from $g_1$) and \ref{fig:g2Mev} (pseudodata from $g_2$). 

The formula for the evidence is given in Eq.~(\ref{eq:Msigpost2}). 
Here $\chi_{aug,min}^2$ decreases as the polynomial degree increases, as adding another parameter to the model results in a fit that is at least as good as that of the previous model. This exponential term increases the evidence for larger values of $M$, but flattens out once we reach the situation that adding another term to the polynomial does not improve the fit to (pseudo)data. Once this situation occurs, increasing $M$ by 1 means that the first term acquires another factor of $1/\sigma_a$. The inverse square root of $\det(A_{aug})$ has nearly the opposite effect as $M$ increases, as adding another row to the matrix corresponds to introducing another uncorrelated parameter with variance $\sigma_a^2$. This explains why the posterior flattens out at large values of $M$, cf. Ref.~\cite{Wesolowski:2016}. 

In Fig.~\ref{fig:g1Mev} we do not see the evidence plot flattening out before we reach $M=6$. In contrast, $g_2$ has an obvious evidence peak, at $M=1$. In the case of $g_1$ the evidence eventually has its highest value around $M \approx 15$---beyond the domain of Fig. ~\ref{fig:g1Mev}. This is not truly a peak for a superior model, but rather a random fluctuation in an essentially flat $\pr(M|D)$ curve.

This means that we have to make a choice for the maximum value of $M$ we will consider when forming our Bayesian Model Average. We call this value $M_{max}$ and it defines 
the set of models used in averaging.

Averaging too few models, for instance stopping at $M_{max}=3$, does not adequately capture the behavior of higher-$M$ models. Averaging too many models, e.g., extending to $M_{max}=100$, overemphasizes the behavior of these high-$M$ models. The observable mean for each non-mixed model $\overline{f}_M(x)$ converges quite quickly to a set value as $M$ increases, and the evidence $\pr(M|D)$ also becomes near-constant at the same limit.  Including too many similar large-$M$ models in the Model Average will result in them dominating the BMA posterior.

Indeed, the BMA procedure assumes that the different non-mixed models provide independent information and posteriors to the mixed model, but the fact that evidence and mean converge for high $M$ indicates that at some value of $M$ model estimates are no longer independent. 
Stopping at $M_{max}=6$ allows us to capture the full diversity of the available models without too much repetition of non-independent models in our model set. We consider models from $M=0$ to $M=6$ in all subsequent results presented here.

\subsection{Observable distributions for different extrapolation distances}

\begin{figure}[th]
\begin{subfloat}[]
{    \includegraphics[width=.5\textwidth]{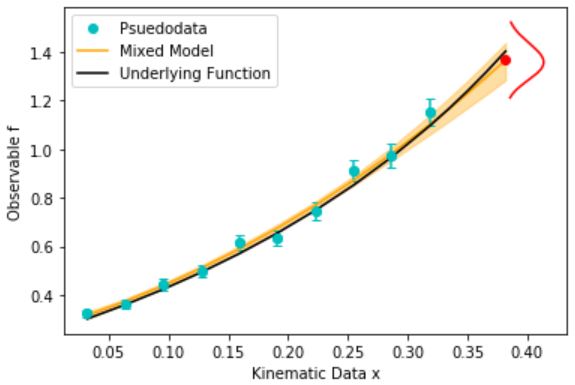}
  \label{fig:g1ex1}}
\end{subfloat}%
\begin{subfloat}[]
    {\includegraphics[width=0.5\textwidth]{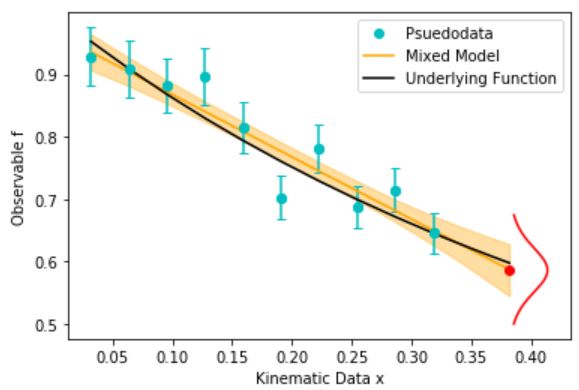} 
  \label{fig:g2ex1}}
\end{subfloat}
\begin{subfloat}[]
    {\includegraphics[width=0.5\textwidth]{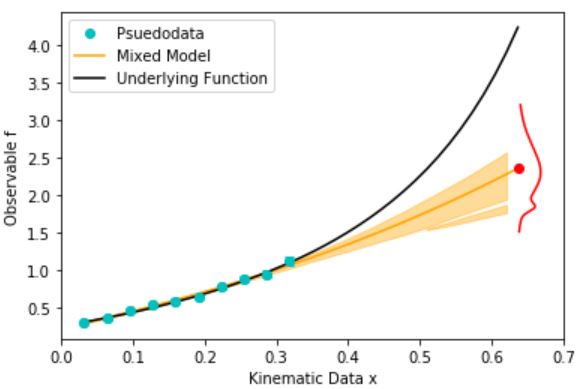}
  \label{fig:g1ex2}}
  \end{subfloat}%
\begin{subfloat}[]
    {\includegraphics[width=0.5\textwidth]{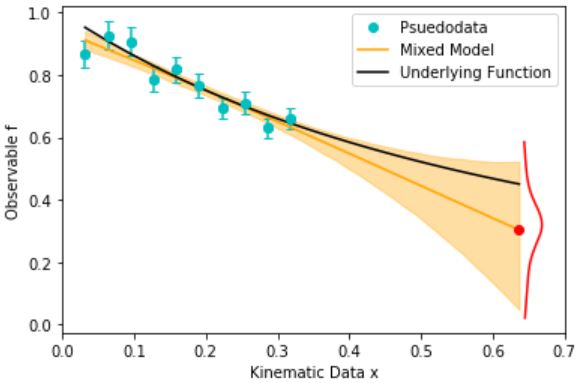}
  \label{fig:g2ex2}}
\end{subfloat}
\caption{Extrapolations for $g_1$ (left column) and $g_2$ (right column), to target points $x_t=1.2/\pi$ (top row) and $x_t=2/\pi$ (bottom row).
In each case $M_{max}=6$ and the bands represent the 68\% CI.
Extrapolation from $g_1$ is difficult, as this underlying function diverges, and so the short distance extrapolation has an underestimation bias which is only accentuated by larger distance extrapolation. In contrast, $g_2$ is easier to extrapolate, and the most consistent issue we see is that posterior width increases for larger distance extrapolations. 
The bands represent ${\rm CI}[0.68]$, so this region encompasses 68\% of the area under the posterior distribution, although note that it may be the union of disjoint intervals, as happens in panel (c).}
\label{fig:test1}
\end{figure}

After determining the observable function posterior $\pr(f|D,M)$ and the mixed posterior $\pr(f|D)$, we wish to test how well these extrapolations perform compared to data to a target point $x_t$. Our pseudodata always cover the domain $x \in [0,1/\pi]$. We have tested individual target points ranging from $x_t=1.2/\pi$ to $x_t=2/\pi$, although only results for these two extremes are discussed below. Fig.~\ref{fig:test1} shows extrapolations for $x_t=1.2/\pi$ and $x=2/\pi$, for $g_1$ and $g_2$, where each set of error bands encloses 68\% of the probability for $\pr(f(x)|D)$.

These 68\% error bands are one set of Credibility Intervals (CIs) on the pdf: they enclose the smallest region of $f$ that contains 68\% of the distribution. 
In general we define ${\rm CI}[\alpha]$ as the smallest interval around the maximum of the pdf that encloses $100*\alpha\%$ probability (``Highest Probability Density'' interval). 
Fig. ~\ref{fig:g1ex2} demonstrates that CIs can be the union of disjoint intervals, though this only occurs for multimodal distributions $\pr(f|D)$.

To quantify the spread of $f$, we take a set of CIs ${\rm CI}[\alpha]$ on the posterior $\pr(f(x_t)|D)$, which we will use in Section ~\ref{sec:cid}. We use $\alpha=[0.1974, 0.383, 0.5468, 0.6827, 0.866, 0.954, 0.987]$.
If we were to calculate these CI[$\alpha$] on a Gaussian distribution (for example $\pr(f|D,M,\sigma_a)$), they would correspond to neat intervals of $[\pm0.25\sigma_a$, $\pm0.5\sigma_a$, $\pm0.75\sigma_a$, $\pm1\sigma_a$, $\pm1.5\sigma_a$, $\pm2\sigma_a$, $\pm2.5\sigma_a]$. Our marginalized and mixed observables $\pr(f|D,M)$ and $\pr(f|D)$ are not Gaussian, but rather a sum of Gaussians, but these CIs can still be defined for each pdf.

Short-distance extrapolations to $x=1.2/\pi$ are usually successful. The distance from the data to the target point is short, and our finite polynomial models (along with the associated mixed model) will typically follow the underlying distribution well. There is a slight difference in the predictive performance of $f$ between $g_1$ in Fig.~\ref{fig:g1ex1} and $g_2$ in Fig.~\ref{fig:g2ex1} at this distance, which is accentuated at larger extrapolation distances.

At large distances, $f$ performs very poorly on $g_1$ (see Fig.~\ref{fig:g1ex2}) and does well on $g_2$ (see Fig.~\ref{fig:g2ex2})---at least for this specific pseudodata set. Though the $g_2$ mean often matches the underlying function, the error bars become very large. This is unavoidable, as the model uncertainty and LEC uncertainty are large: the low-$x$ data is more distant from the target point $x_t$ and thus does not constrain the observable precisely at $x_t=2/\pi$. The $g_1$ underlying function diverges beyond any finite polynomial as the target point gets farther from the data, and even our mixed model cannot match the function at this distance. The underlying function value falls a large distance from the observable mean, in the distant tails of the observable posterior, which the pdf predicts with very low probability.

We therefore reach the qualitative conclusion that BMA seems to extrapolate quite well for some situations (short distances, $g_2$ at more distant target points), but does not produce a good model in all situations ($g_1$). In particular, we have demonstrated that BMA only yields a useful statistical model if its constituent set of models contains models that represent the underlying function well. This means, of course, that our toy problems are particular examples of general theorems regarding the convergence of BMA in the ${\cal M}$-open setting~\cite{Chib:2016}. 

\section{Assessing predictive performance using the Credibility Interval Diagnostic}
\label{sec:cid}

\subsection{Defining and Calculating the Credibility Interval Diagnostic} \label{ssec:ciddefine}

Since we are working with pseudodata generated from a known function, we are able to determine the accuracy of a particular EFT extrapolant by comparing its pdf at a given target point $x_{t}$ to the data generated from the underlying function $g$. We start by generating a validation datum $d_t$ at $x_t$ from the Gaussian distribution around $g(x_t)$; the width of this Gaussian is defined by the same fractional error we took for the small-$x$ pseudodata sets. We will compare this validation data point to different model extrapolations $\pr(f(x_t)|D,M)$ and $\pr(f(x_t)|D)$, thus identifying how well our models predict where new `experimental' data occurs at kinematic points above our low-$x$ data sets.

After specifying this data point $d_t$ we hold it constant while generating and fitting our models to a number of pseudodata sets. The extrapolants $f$ change based on the polynomial order we use and the pseudodata we fit to. We compute how the predicted values $f(x_t)$ compare to the validation datum $d_t$ for a large number of pseudodata sets. This provides an assessment of the performance of the extrapolant due to uncertainties in the low-$x$ data.
After we have compared several $f$'s to this single $d_t$ we discard it and draw from the Gaussian distribution around $g(x_t)$ again to generate a new validation datum. We then hold this new $d_t$ constant while again varying the pseudodata and generating a new extrapolant from each pseudodata set. By doing repeated draws from the distribution around $g(x_t)$ we can see how much the randomness in the target value affects our assessment of the extrapolants' performance.

As the distribution of $f$ changes with different pseudodata, a single validation data point $d_t$ ``falls'' in different portions of our model pdf. Sometimes the pdf density 
will be large at $d_t$, and sometimes it will be small. When evaluated over a large number of data sets a model with accurate predictions will frequently produce  a $f(x_t)$ posterior with high probability at $d_t$. 
To compare the validation data to the model prediction $f$, we use a predictive performance metric previously employed in the Bayesian nuclear-physics context in Refs.~\cite{Furnstahl:2015,Melendez:2017,Neufcourt:2018,Melendez:2019}. Consider the credibility intervals ${\rm CI}[\alpha]$ defined above on the distributions $\pr(f|D)$ and $\pr(f|D,M)$. 
We want to check how often a validation data point falls within the CIs defined for each model. 
For instance, we want to determine if $d_t$ falls inside ${\rm CI}[0.50]$ for 50\% of the different data sets used to form the model prediction. A model has good predictive performance if it  accurately encodes the probability of finding the validation datum $d_t$ in a given CI of the extrapolated $f$.

To quantify the predictive performance of either the BMA result or a fixed-degree polynomial we keep $d_t$ constant, generate $N$ random data sets, $D_i$, $i=1,\ldots,N$, and fit the model in question to each of them.
This generates $N$ extrapolants to the target point $x_t$ in each model. For each data set $D_i$, we find the CIs of $\pr(f(x_t)|D_i)$ [or $\pr(f(x_t)|D_i,M)$] and define ${\rm CI}_{i,x_t}(\alpha)$ as the $100*\alpha$\% interval of the $f$ posterior for that $i$th data set. 
The proportion of data sets for which the validation data point $d_t$ lies within a given Credibility Interval in the model is
\begin{equation} 
\label{eq:indicatorfinal}
D_{x_{t}}(\alpha) \equiv \frac{1}{N}\sum_{i=1}^{N}\mathbf{1}[d_t\in {\rm CI}_{i,x_t}[\alpha]],
\end{equation}
where $\mathbf{1}$ is the indicator function, so the term being summed is 1 if the validation data is within the CI and 0 otherwise. This counts ``hits and misses'' on the credibility intervals: 
$D_{x_t}(\alpha)$ measures the frequency of hits at $x_t$ in a model's CI$[\alpha]$. By comparing $D_{x_{t}}(\alpha)$ to $\alpha$ we determine how the area under the pdf compares to the frequency with which a validation datum falls within CI[$\alpha$].
This diagnostic originates from Ref.~\cite{Bastos:2009}, and this particular formulation is from Ref.~\cite{Melendez:2019}. The results below are generated using $N=100$ data sets $D_i$. 

An accurate statistical model that describes the uncertainties perfectly would have a ${\rm CI}[\alpha]$ within which $d_t$ falls $100*\alpha$\% of the time, i.e., for a perfect model, $D_{x_{t}}(\alpha)=\alpha$. If a model produces values for $D_{x_t}(\alpha)$ that fall far from this ideal line that indicates that it predicts poorly.
We note that such an interpretation of the model's predictive performance does assume independent trials; that assumption is not always satisfied.

If the measurement at $g(x_t)$ had no uncertainty then $D_{x_t}(\alpha)$ would not vary upon repeated draws from the distribution of possible measurement outcomes. 
In this case the validation datum has no variability and either always falls into a particular CI$[\alpha]$ or never falls in that interval. But, under the circumstances of our toy-problem test, a range of outcomes for the value $d_t$ will be obtained upon repeated sampling of the measurement at $x_t$. The movement of $d_t$ will change the number of hits and misses on the right-hand side of Eq.~(\ref{eq:indicatorfinal}), so these data fluctuations at $x_t$  make $D_{x_{t}}(\alpha)$ a distribution rather than a function. 
To assess the width of this distribution, after finding one line $D_{x_{t}}(\alpha)$ vs $\alpha$ for each fixed-$M$ model and for the BMA model, we repeat the process and find the function $D_{x_{t}}(\alpha)$ for a different validation data point $d_t$ drawn from the same Gaussian distribution around $g(x_t)$. We repeat the computation of $D_{x_t}(\alpha)$ as a function of $\alpha$ 20 times overall so that we have 20 $D_{x_{t}}(\alpha)$ lines for each extrapolant and each $x_t$. 
We represent the resulting distribution of $D_{x_t}(\alpha)$ lines as a band to show the distribution of results for the Credibility Interval Diagnostic. The quantities $D_{x_t}(\alpha)$ are also sometimes called ``Empirical Coverage Probabilities", cf. \textcite{Neufcourt:2018} and references therein.
This band includes the central 70\% (14 out of our 20) of the  lines generated for different $d_t$'s.  The central line within the band indicates average performance of each model as regards extrapolation to $x_t$ from different pseudodata sets, and the width of the band comes from variation in validation data. 

\subsection{Credibility Interval Diagnostic results for different extrapolation distances}
\label{sec:predperf}

\begin{figure} [h]
\begin{subfloat}[]
    {\includegraphics[width=0.48\textwidth]{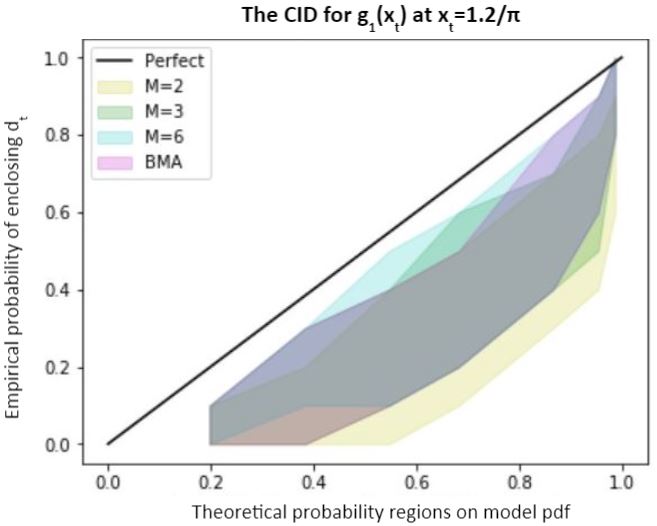}
    \label{fig:g1cid1}}
\end{subfloat}
\begin{subfloat}[]
    {\includegraphics[width=0.48\textwidth]{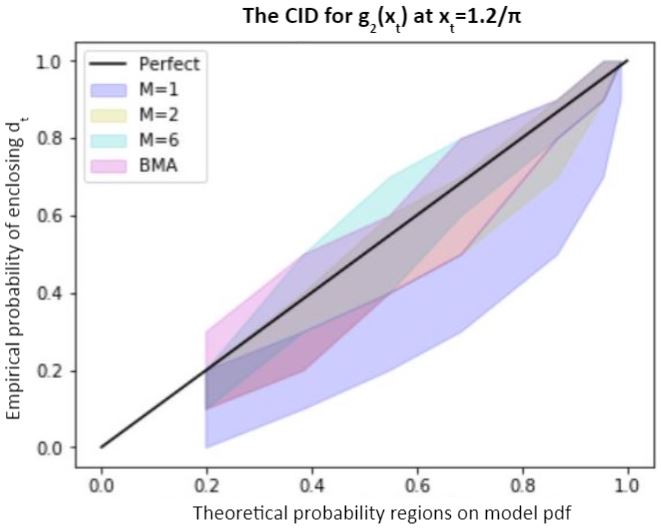}
    \label{fig:g2cid1}}
\end{subfloat}
\begin{subfloat}[]
{\includegraphics[width=0.48\textwidth]{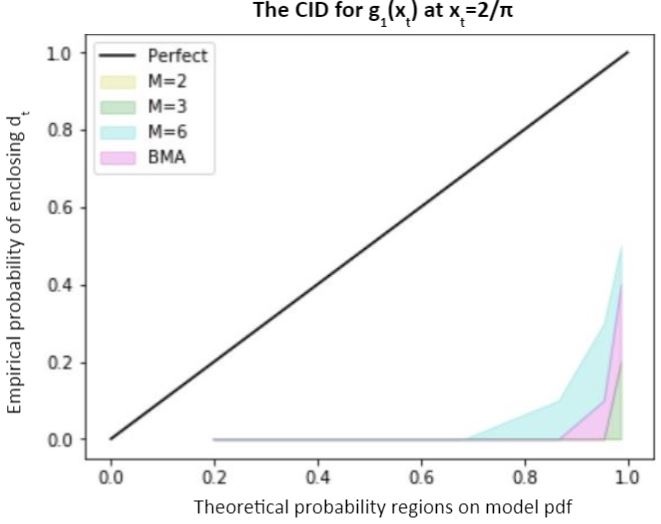}
    \label{fig:g1cid2}}
  \end{subfloat}%
\begin{subfloat}[]
{\includegraphics[width=0.48\textwidth]{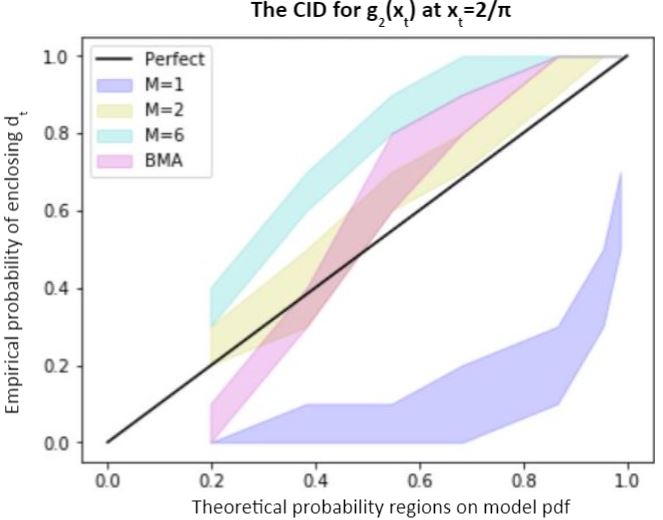}
    \label{fig:g2cid2}}
\end{subfloat}
 \caption{Credibility Interval Diagnostic (CID) plots for four different extrapolations considered in this paper. Panels (a) and (c) show results for a nearby ($x_t=1.2/\pi$) and distant ($x_t=2/\pi$) extrapolation of the function $g_1$. Panels (b) and (d) show CIDs at the same target points for the function $g_2$. In each panel the bands that are not pink show the performance of extrapolations for specific polynomial orders according to the legend, while the pink band is the predictive performance of the BMA extrapolant.}
\label{fig:test2}
\end{figure}

The resulting plot is the Credibility Interval Diagnostic, see Figs.~\ref{fig:g1cid1}-- \ref{fig:g2cid2}. It evaluates whether a model has CIs that describe  the certainty of its predictions well. In Figs.~\ref{fig:g1cid1}--\ref{fig:g2cid2} we show the CID for $g_1$ and $g_2$ at two different target points, $x_t=1.2/\pi$ and $x_t=2/\pi$ so we can assess how well each EFT expansion extrapolates from the pseudodata it models.

Our Credibility Interval Diagnostic plots show results for different polynomial orders fit to pseudodata from $g_1$ and $g_2$ as well as results for the BMA extrapolant. For $g_1$ we show $M=2$, the lowest-$M$ model that has nonzero $D_{x_{t}}(\alpha)$ on the CID; $M=6$ as our highest-$M$ model (this also has the highest evidence for our non-mixed models); and $M=3$ so there is some indication of how a model between $M=2$ and $M=6$ performs. For $g_2$ we show the highest-evidence model ($M=1$), the nearest-to-ideal prediction ($M=2$),  and the highest order ($M=6$). All plots also show the mixed model ($M_{max}=6$). 

In each panel $D_{x_{t}}(\alpha)=100*\alpha$\% (the line in black) defines perfect prediction. 
If a model band lies underneath this  line its posterior CIs are too small and/or misplaced: the CIs do not encompass the validation data as often as their stated probability $\alpha$ would indicate. If a model lies above this line, its posterior is too wide: the uncertainty is so large that a given CI receives more validation data `hits' than it should. 
A model that crosses or straddles the line has both issues, depending on the value of $\alpha$.

For $g_1$ we see that the mixed model acts like the fixed-order models that comprise it. 
This analysis of CIDs for data from $g_1$ shows that in this case BMA fails to extrapolate from the low-$x$ data. At both $x_t=1.2/\pi$ (Fig.~\ref{fig:g1cid1}) and $x_t=2/\pi$ (Fig.~\ref{fig:g1cid2}), the non-mixed models fail to predict well: the CIs are not centered around the underlying function and are too small to encompass $d_t$. The models all fall below the perfect prediction line, with the gap widening as the extrapolation distance increases. The mixed model sits somewhere in the middle of the fixed-degree models, and thus also has poorly-located and overconfident CIs. 

$g_2$ gives us a better understanding of how BMA succeeds when we have useful models. 
Recall from Fig.~\ref{fig:g2Mev} that $M=1$ has the highest evidence for a non-mixed model in $g_2$, but it has much too small CIs---its pdf has too thin a peak. It therefore falls under the perfect prediction curve on the CID plot despite its high evidence.
$M=2$ follows the perfect prediction line well; its credibility intervals are very nearly the actual probability of verification data falling within them.
Meanwhile, the $\chi_{aug}^2$ is smallest for the $M=6$ model. It  ``fits" the data best, but it has CIs that are too large---much too large for $x_t=2/\pi$. The higher-order LECs are poorly defined by the low-$x$ data and return the prior, producing too large an uncertainty at the extrapolation point. 

The BMA prediction obtained with data from $g_2$ crosses the line of perfect prediction. Here very small values of $\alpha$ have too small ${\rm CI}[\alpha]$, while values near $\alpha=1.0$ have overly-large ${\rm CI}[\alpha]$. 

We have found this behavior is somewhat generic: even when mixed models are moderately successful predictors they still tend to exhibit this ``S-shape''
 CI behavior.
The S-shape is a reflection of the BMA posterior pdfs, which have high peaks corresponding to the highest-evidence, small-$M$ polynomial pfs, and long tails from their lower-evidence, larger-$M$ polynomial pdfs, see, e.g., Figs.~\ref{fig:g1BMA} and \ref{fig:g2BMA}.

For small values of $\alpha$, our CIs are small intervals around the highest peaks of $\pr(f|D)$, and the mixed model $D_{x_{t}}(\alpha)$ inherits the predictive performance of the highest-evidence models included in the average. In the case of $g_2$ that model is $M=1$, and the too-narrow CIs of that model yields a mixed-model $D_{x_{t}}(\alpha)$ that lies under our perfect prediction line. 

For values of $\alpha$ near 1, the tails of the longest-tailed models ($M=6$ in this case) dominate the mixed-model posterior; to encompass 99\% of the posterior probability, the CIs must extend far along these tails. This makes the mixed-model CIs too large near $\alpha=1$. 
For the prediction of $g_2(2/\pi)$ the large tails associated with $M=3$ through $M=6$ give mixed-model CI$[\alpha]$'s that cover 100\% of the validation data already for $\alpha \approx 0.8$.

For sufficiently large extrapolation distances the mixed model has elements of both overly narrow and overly broad models, and its predictive performance suffers accordingly. 
\section{Conclusions}
\label{sec:end}

The toy problems we examined here are designed such that extrapolating from low-$x$ data is bound to eventually fail. Polynomial extrapolations to points outside the radius of convergence of the Taylor series should not succeed. For the first function, $g_1$, extrapolations to target  points inside, but near, the radius of convergence inherit this failure mode, since the $\tan$ function diverges at $x=1$, so every finite polynomial model falls underneath the function itself. 
We tried polynomial models with degree up to $M=100$ with $\sigma_a=5$, and the validation data barely falls within the $1\sigma$ range of our Gaussian $f_{M=100}(x_t)$. There is no low-degree polynomial that will successfully extrapolate from a data set generated from $g_1$. Mixing polynomials does not remedy that problem. BMA simply is not useful unless the full set of models has useful members.

A contrasting case is the extrapolation problem for the function $g_2$. This function is sufficiently docile that BMA is advantageous. Even though approximating $g_1$ and $g_2$ by polynomials both constitute ${\cal M}$-open situations, the practical performance of BMA in the two cases is very different. In the case of $g_2$ neither $M=1$ or $M=6$ has good predictive performance: neither high evidence or smallest $\chi^2$ reliably indicates that a model predicts well. 
The model $M=2$ performs superbly well.
This is easy to diagnose in a toy problem, because the underlying function's value is calculable at the target point. If this were not a toy problem, there would be no computable metric to tell us $M=2$ had reputable CIs. In a real extrapolation problem the experiments taken in our limited range are all the data we have available.
The BMA prediction for $g_2(x_t)$ is worse than $M=2$, but better than both the highest evidence ($M=1$) and highest order  ($M=6$) model. BMA gives us a model that outperforms a naively selected model $M$, but it may not perform as well as a single model that happens to extrapolate accurately.  It does, though, have better forecasting ability better than most specific-$M$
models. 

If there is a ``correct'' model describing the data, this model is incorporated into the mixed model, but its prediction is diluted by other models that have relatively high evidences. The mixed model will still have elements of this ``correct'' model, but will likely have worse predictive performance than a single very well-performing model ($M=2$ for $g_2$, for example). The benefit of BMA is that the mixed model has better predictive performance than \textit{most} non-mixed models. Choosing the mixed model will most often present a better forecast than selecting an arbitrary non-mixed model, and this can be useful when there is significant model uncertainty. We prevent ourselves from being completely wrong by refusing to choose one non-mixed model, at the cost of degraded performance in comparison to the case where we are lucky enough to guess the correct answer.


Indeed, evidence is an incomplete metric for model appropriateness when extrapolating, since it is computed using data in an area and assumes the model is equally applicable at the extrapolation point. Evidence also is sensitive to the whole posterior, including parts that are not very well determined by data~\cite{Gelman:2020oco}. There are other measures of predictive performance that might better select or better weight the models, and these ought to be explored in future research on combining model posteriors for extrapolation~\cite{Gelman03,Yao:2018}. 

Some aspects of our statistical approach could be improved. The toy problems
do not yield Taylor series with random coefficients. The LEC prior assumes that positive and negative coefficients will come up in equal proportion, since they are independent draws from a ``naturalness distribution'' centered at zero. The toys, however, were chosen to have specific patterns in their LECs: they are all positive in $g_1$ and alternate between positive and negative in $g_2$. Since our priors on the LECs do not account for these correlations, the analysis is ignorant of this pattern, and the extrapolation suffers as a result. A topic for future work is to develop better statistical models that learn, and then use, the correlation pattern between coefficients to improve the extrapolation~\cite{Bonvini:2020xeo}.



\appendix
\section{Computing $\pr(\sigma_a|D,M)$}
\label{appendix1}
The posterior $\pr(\vec{a}|D,M,\sigma_a)$ can be written in terms of an augmented $\chi^2$, $\chi^2_{aug}$, see Eq.~(\ref{eq:LECpost3}). We follow Ref.~\cite{Schindler:2009} and write $\chi_{aug}^2$ in matrix form  
\begin{equation}
\chi_{aug}^2(\vec{a})=\vec{a}^T\matA_{aug}\vec{a}-2\vec{b}\vec{a}+C=(\vec{a}-\vec{a}_0)^T\matA_{aug}(\vec{a}-\vec{a}_0)+\chi_{aug,min}^2,
\end{equation}
where
\begin{equation} \label{eq:twoparts}
\chi^2_{aug}=\chi^2+\vec{a}^T\frac{1}{\sigma_a^2}\vec{a}, 
\end{equation}
with the standard $\chi^2$ given by
\begin{equation}
\chi^2=\vec{a}^T\matA\vec{a}-2\vec{b}\vec{a}+C.
\end{equation}

The minimum $\chi^2_{aug}$ then occurs at parameter values
\begin{equation}
\vec{a}_{0}=\matA^{-1}_{aug}\vec{b}
\end{equation}
and is
\begin{equation}
\chi^2_{aug,min}=\chi^2_{aug}(\vec{a}_{0})=C-\vec{b}^T \matA_{aug}\vec{b}.
\end{equation}
The minimizing $\vec{a}$ for the original, non-augmented $\chi^2$, which we denote by $\vec{\alpha}_0$, is then:
\begin{equation}
\vec{\alpha}_{0} \equiv \matA^{-1}\vec{b},
\end{equation}
such that
\begin{equation}
    \chi^2=(\vec{a}-\vec{\alpha}_0)^T\matA(\vec{a}-\vec{\alpha}_0)+\chi^2_{min},
\end{equation}
with
\begin{equation}
\chi^2_{min}=\chi^2(\vec{\alpha}_{0})=C-\vec{b}^T \matA\vec{b}.
\end{equation}

With this vocabulary, we can derive an analytic understanding of the posteriors $\pr(D|M,\sigma_a)$ and $\pr(\sigma_a|D,M)$.
To do this we first note that:
\begin{equation}
\vec{a}_{0}=(\matA+\sigma_a^{-2}\mathbb{I})^{-1}\vec{b}=(\matA+\sigma_a^{-2}\mathbb{I})^{-1}\matA\vec{\alpha_0}=(\matA^{-1}+\sigma_a^{2}\mathbb{I})^{-1}\sigma_a^2\vec{\alpha_0},
\end{equation}
which allows us to write
\begin{equation}
\vec{\alpha_0}-\vec{a}_{0}
=\left[\mathbb{I}-(\matA^{-1}+\sigma_a^{2}\mathbb{I})^{-1}\sigma_a^2 \right]\vec{\alpha_0}=(\sigma_a^2\matA+\mathbb{I})^{-1} \vec{\alpha_0}.
\label{eq:diff}
\end{equation}

Now we begin to write $\chi^2_{aug,min}$ in terms of both minimal vectors $\vec{a}_0$ and $\vec{\alpha}_0$ using Eq.~(\ref{eq:diff}):
\begin{equation}
\chi^2_{aug,min}=\vec{a}_0^T \frac{1}{\sigma_a^2} \vec{a}_0 + (\vec{a}_0-\vec{\alpha}_0)^T\matA(\vec{a}_0-\vec{\alpha}_0)+\chi^2_{min},
\end{equation}
which means that
\begin{equation}
\chi^2_{aug,min}= \vec{\alpha_0}^T((\sigma_a^2\matA+\mathbb{I})^{-1})^T\matA(\sigma_a^2\matA+\mathbb{I})^{-1}\vec{\alpha_0}+\vec{\alpha_0}^T\left[(\matA^{-1}+\sigma_a^{2}\mathbb{I})^{-1}\right]^T\sigma_a^2(\matA^{-1}+\sigma_a^{2}\mathbb{I})^{-1} \vec{\alpha_0}+\chi^2_{min}.
\end{equation}
The can be simplified to yield the result:
\begin{equation}
\chi^2_{aug,min}=\vec{\alpha_0}^T(\sigma_a^2 \mathbb{I} +\matA^{-1})^{-1} \vec{\alpha_0}+\chi^2_{min}
\label{eq:chisqaugmin}
\end{equation}

In the limit where $\sigma_a^2 \gg \sigma_k^2$, i.e., the pseudodata has much smaller errors than the prior range for the LECs, this simplifies further to
\begin{equation}
\chi^2_{aug,min}=\frac{\vec{\alpha_0}^T\vec{\alpha_0}}{\sigma_a^2}+\chi^2_{min},
\end{equation}
i.e., the main effect of the prior is to increase the overall value of $\chi^2_{aug,min}$ over $\chi^2_{min}$ and there is no change in the position of the minimum to leading order.

This sort of analysis also helps us find the posterior $\pr(\sigma_a|D,M)$ as does our use of a conjugate, inverse $\chi^2$ prior on $\sigma_a^2$. Recall that $\pr(\sigma_a)$ is independent of $M$.
Our posterior will be
\begin{equation}
    \pr(\sigma_a|D,M) \propto \pr(D|\sigma_a,M)\pr(\sigma_a|M)
=\int\pr(D|\vec{a},M,\sigma_a)\pr(\vec{a}|M,\sigma_a)\pr(\sigma_a) d \vec{a}.
\end{equation}
Evaluation of the Gaussian integral using the Laplace approximation---which is exact in this case---yields
\begin{equation}
    \pr(\sigma_a|D,M)=\frac{1}{\sigma_a^{M+1}}\frac{1}{\sqrt{\det(A_{aug})}}\exp\left(-\frac{\chi^2_{aug,min}}{2}\right)\pr(\sigma_a).
\end{equation}

Using (\ref{eq:chisqaugmin}) this becomes:
\begin{equation}
    \pr(\sigma_a|D,M)=\frac{1}{\sigma_a^{M+1}}\frac{1}{\sqrt{\det(A+\sigma_a^{-2}\mathbb{I})}}\exp\left(-\frac{\vec{\alpha_0}^T(\sigma_a^2 \mathbb{I} +\matA^{-1})^{-1} \vec{\alpha_0}+\chi^2_{min}}{2}\right)\pr(\sigma_a)
    \label{eq:sigmapost}
\end{equation}
Now we diagonalize our $A$, defining
\begin{equation}
\matA=O^T\Delta O,
\end{equation}
where $\Delta$ contains the eigenvalues
of $\matA$, $\Delta_i$, $i=0,1,2\ldots M$.
This allows us to compute the determinant in question as
\begin{equation}
    \det(A+\sigma_a^{-2}\mathbb{I})=\det(O^T\Delta O+\sigma_a^{-2}\mathbb{I} )=\det(\Delta+\sigma_a^{-2}OO^T)
    =\prod_{i=0}^M (\Delta_i+\sigma_a^{-2}).
    \label{eq:det}
\end{equation}

We then perform a similar decomposition for $\chi^2_{aug,min}$.
\begin{equation}
    \left(\vec{\alpha_0}^T \left(\sigma_a^2 \mathbb{I} +\matA^{-1}\right)^{-1} \vec{\alpha_0}+\chi^2_{min}\right)=\left(\left(\vec{\alpha_0}O\right)^T \left(\sigma_a^2 \mathbb{I} +\Delta^{-1}\right)^{-1}  \left(O\vec{\alpha_0}\right)+\chi^2_{min}\right).
    \end{equation}
This renders computation of the exponential term in Eq.~(\ref{eq:sigmapost}) straightforward
\begin{equation}
    \exp\left(-\frac{\chi^2_{aug,min}}{2}\right)=  \exp\left(-\frac{\chi^2_{min}}{2}\right) \prod_{i=0}^M \left(\exp\left(-\frac{1}{2} \frac{(\sum_j O_{i,j} a_i)^2}{\Delta_i^{-1}+ \sigma_a^2}\right)\right).
    \label{eq:Pstar}
\end{equation}
Substituting Eqs.~(\ref{eq:Pstar}) and (\ref{eq:det}) into Eq.~(\ref{eq:sigmapost}) and combining it with the prior $\pr(\sigma_a)$ isolates all the $\sigma_a$ dependence in the posterior and therefore provides an analytic formula for the behavior observed in Fig.~\ref{fig:test00}.

\acknowledgments{We thank Dick Furnstahl, Witek Nazarewicz, Matt Plumlee, and Matt Pratola for useful discussions and Alexandra Semposki for a careful reading of the manuscript. The work of M.C. and I.B. was supported by Research Apprenticeships from the Honors Tutorial College at Ohio University. The work of D.R.P. was supported by the US Department of Energy under contract DE-FG02-93ER-40756 and by the National Science Foundation CSSI program under award number OAC-2004601 (BAND Collaboration).}


\end{document}